\documentclass[dvips]{elsart}

 \usepackage{epsfig,amsmath}

\begin{document}

 \begin{frontmatter}

   \title{Reaction mechanisms involved in the production of neutron-rich isotopes}

   \author{J.~Benlliure$^1$, K.~Helariutta$^2$, K.-H.~Schmidt$^2$ and M.V.~Ricciardi$^2$}

   \address{$^1$Universidad de Santiago de Compostela, 15706 Santiago de Compostela, Spain}
   \address{$^2$Gesellschaft f\"ur Schwerionenforschung, Planckstrasse 1 \\
64291 Darmstadt, Germany} 


 \begin{abstract}
The reaction mechanisms best suited for the production of neutron-rich nuclei, fragmentation and fission, are discussed. Measurements of the production cross sections of reaction residues together with model calculations allow to conclude about the expected production rates of neutron-rich isotopes in future facilities.
 \end{abstract}
 
 \end{frontmatter}

\section{Introduction}

The production of nuclei far from stability has provided a new ground to investigate the structure and the dynamics of nuclei. The present experimental facilities are able to produce a large variety of neutron-deficient isotopes, however, the neutron-rich side of the chart of the nuclides is only accessible up to the drip lines for the lightest nuclei. Consequently, the challenge for most of the future planned rare-beam facilities is the production of neutron-rich isotopes. 

Together with the technological improvements concerning high-current accelerators, large-acceptance separators and larger extraction and charge-breeding efficiencies, the optimal choice of the reaction mechanisms will play a major role on the final production rates of neutron-rich nuclei. Different reaction mechanisms can be used to produce these nuclei, however, heavy-ion collisions at low energies like fusion, deep inelastic or multinucleon transfer can only be applied with thin targets limiting the final production rates. Better suited seem to be fragmentation or spallation at high energies and fission. In addition, these two reaction mechanisms allow to produce a large variety of neutron-rich nuclei. 

In this paper we discuss the main issues of these two reaction mechanisms. The large collection of data obtained during the last years has brought new model descriptions of these reactions providing reliable predictions on the expected production rates of neutron-rich isotopes in future planned rare-beams facilities.

\begin{figure}[t]
\centerline{\epsfig{file=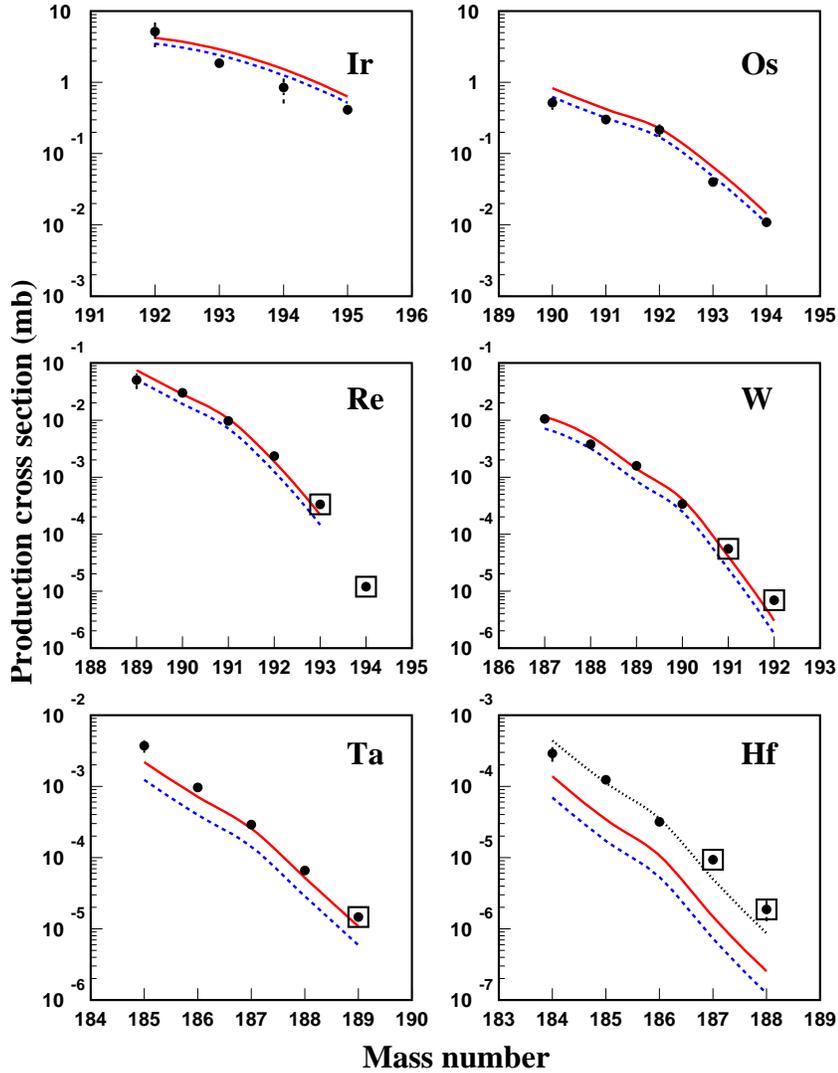,width=11cm}}   
\vspace*{-0.2cm}
\caption{Production cross sections of neutron-rich nuclei in cold-fragmentation reactions of $^{197}$Au at 1 A GeV in a beryllium target. The lines represent different model calculations with the code COFRA (see text). The squares mark those isotopes which were observed for the first time in reference $^6$.} 
\label{fig_1}
\end{figure}

\section{Production of heavy neutron-rich nuclei}

Heavy exotic nuclei (Z$>$70) can be produced by means of fusion-evaporation reactions or in fragmentation(spallation) of heavy nuclei. Both reaction mechanisms lead mainly to the production of neutron-deficient residues. However, it has recently been shown that fragmentation reactions at relativistic energies present large fluctuations in the N/Z distribution of the final residues and in its excitation-energy distribution. In particular, the proton-removal channels have been investigated in cold-fragmentation reactions\cite{ben99} where only protons are abraded from the projectile, while the induced excitation energy is below the particle-emission threshold. These reactions can lead to the production of heavy neutron-rich nuclei beyond the present limit of the chart of the nuclides as shown in Fig. \ref{fig_1}.

Cold-fragmentation reactions can be described in terms of the abrasion-ablation model as a two-step process. First, the interaction between projectile and target leads to a projectile-like residue with a given excitation energy which  statistically de-excitates by particle evaporation or fission. A new analytical formulation of the abrasion-ablation model, the code COFRA,  has been developed\cite{ben99} in order to calculate the expected low production cross sections of extremely neutron-rich nuclei which are not reachable with Monte Carlo codes. 

\begin{figure}[t]
\centerline{\epsfig{file=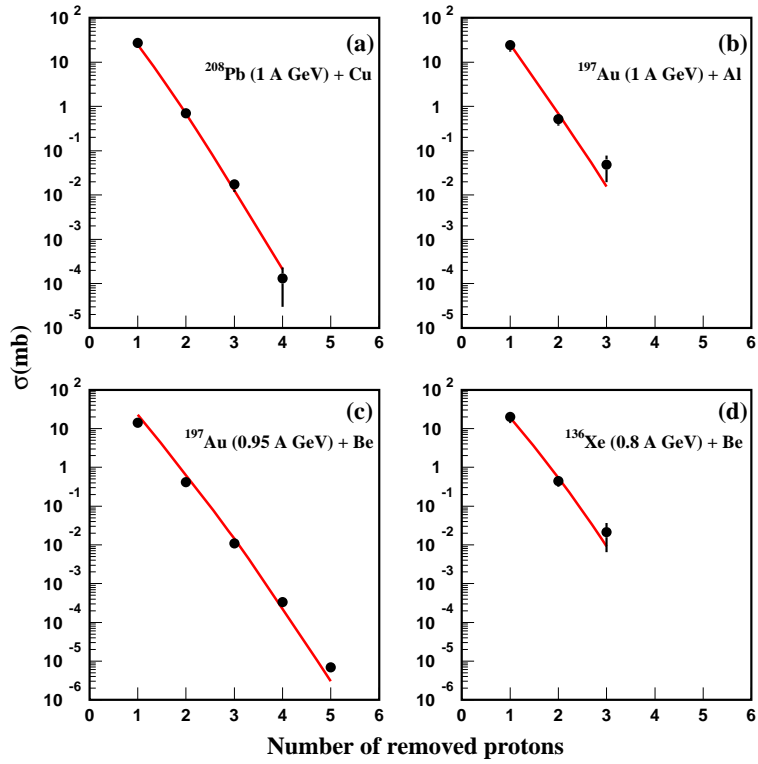,width=11cm}}
\vspace*{-0.1cm}   
\caption{Production cross sections of proton-removal channels in different reactions (dots) $^{2,3,4}$ compared with the predictions of the code COFRA (lines).} 
\label{fig_2}
\end{figure}

The results of these calculations are shown in Fig. \ref{fig_1}. The different lines correspond to calculations with different enhancement factors of the excitation energy induced in the collision due to particle-hole excitations (see reference\cite{ben99} for details). As can be seen, the production cross sections of very neutron-rich nuclei can nicely be described with the new analytical formulation of the abrasion-ablation model. In Fig. \ref{fig_2} we compare the predictions of the code with the production cross sections of proton-removal channels populated in different reactions \cite{ben99,jon98,sch92}. 

\begin{figure}[t]
\centerline{\epsfig{file=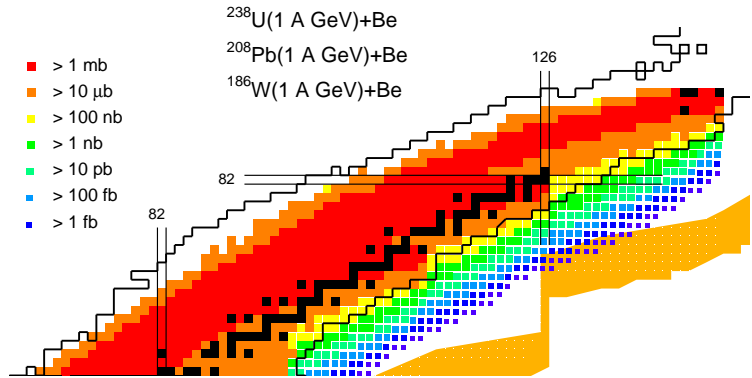,width=11cm}}   
\vspace*{-0.4cm}
\caption{Estimated production of heavy neutron-rich residues in cold-fragmentation reactions induced by $^{238}$U, $^{208}$Pb and $^{174}$W at 1 A GeV in a beryllium target on top of a chart of the nuclides. The color scale indicates the production cross section.} 
\label{fig_3}
\end{figure}

Since the COFRA code describes correctly all the measured proton-removal channels in relativistic heavy-ion collisions, it was used to determine the expected production of heavy neutron-rich nuclei in future rare-beam facilities. The results of these calculations are shown in Fig. \ref{fig_3}. In this figure we report the expected production cross sections of heavy neutron-rich nuclei that can be obtained in the fragmentation of $^{238}$U, $^{208}$Pb and $^{174}$W. According to these calculations, large progress is expected in this region of the chart of the nuclides, where the r-process path may even be reached around the end point N=126.

\begin{figure}[ht]
\hspace*{-0.5cm}
    \epsfig{file=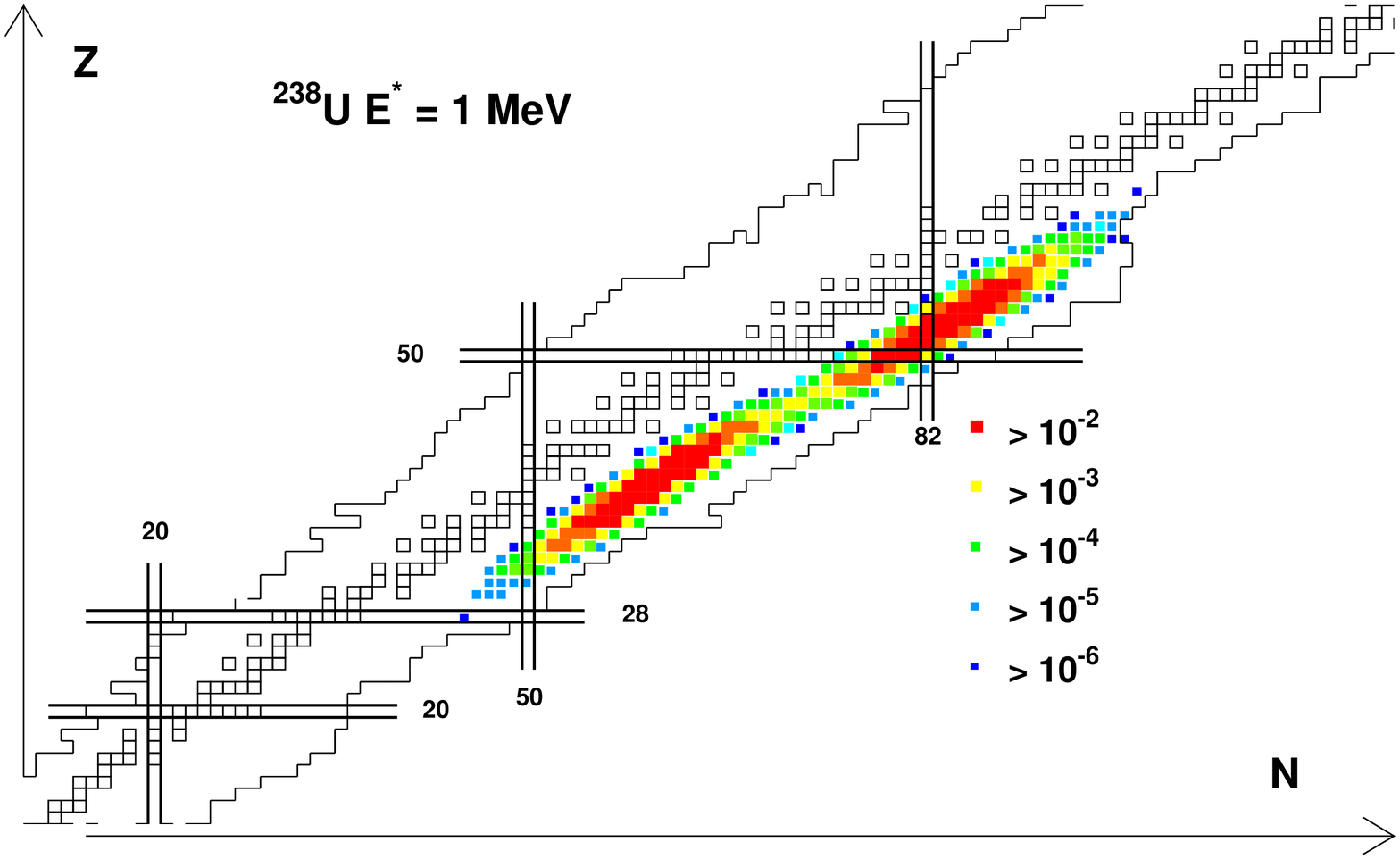,width=7.0cm}
    \epsfig{file=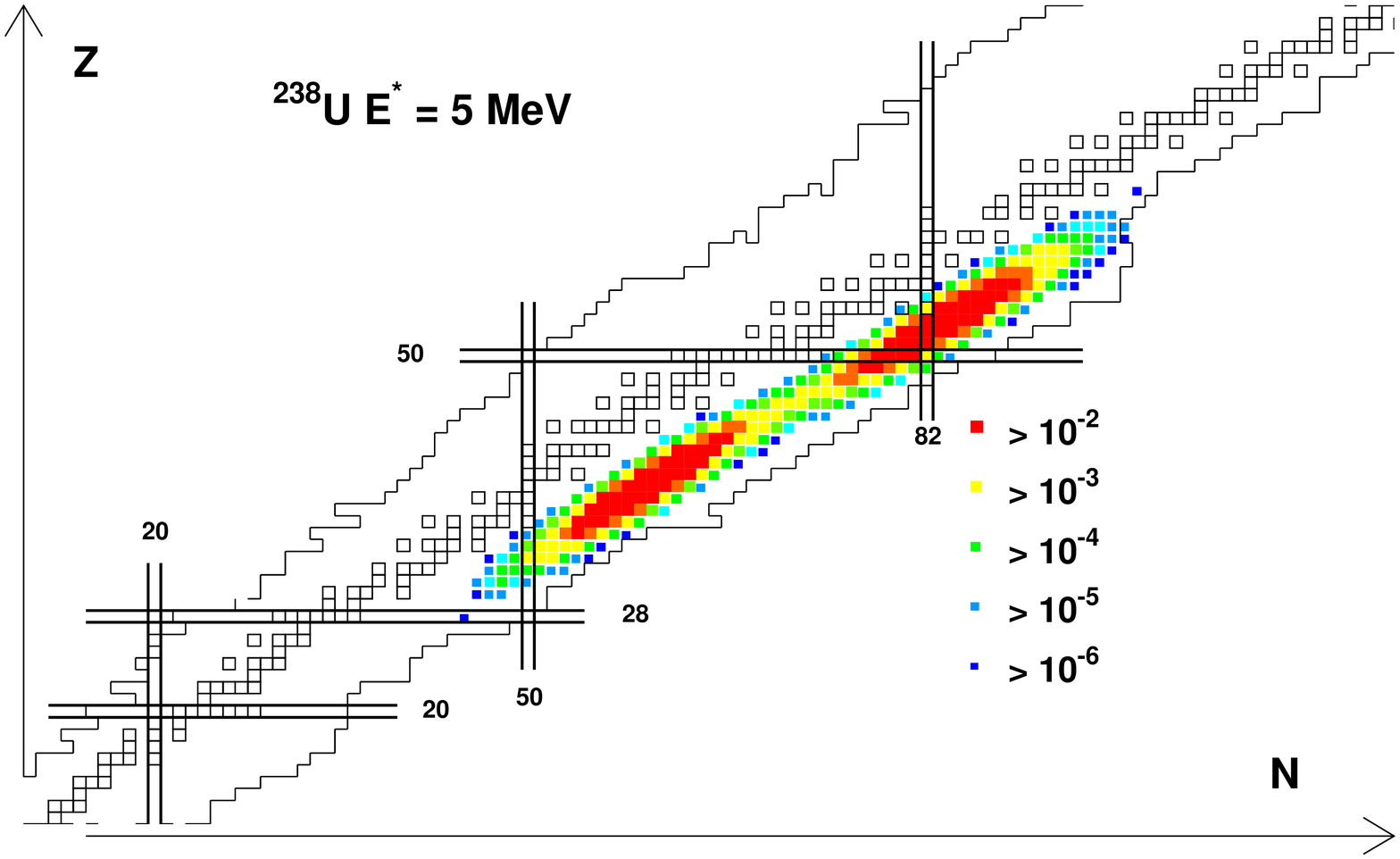,width=7.0cm}

\hspace*{-0.5cm}
\vspace*{0.2cm}
    \epsfig{file=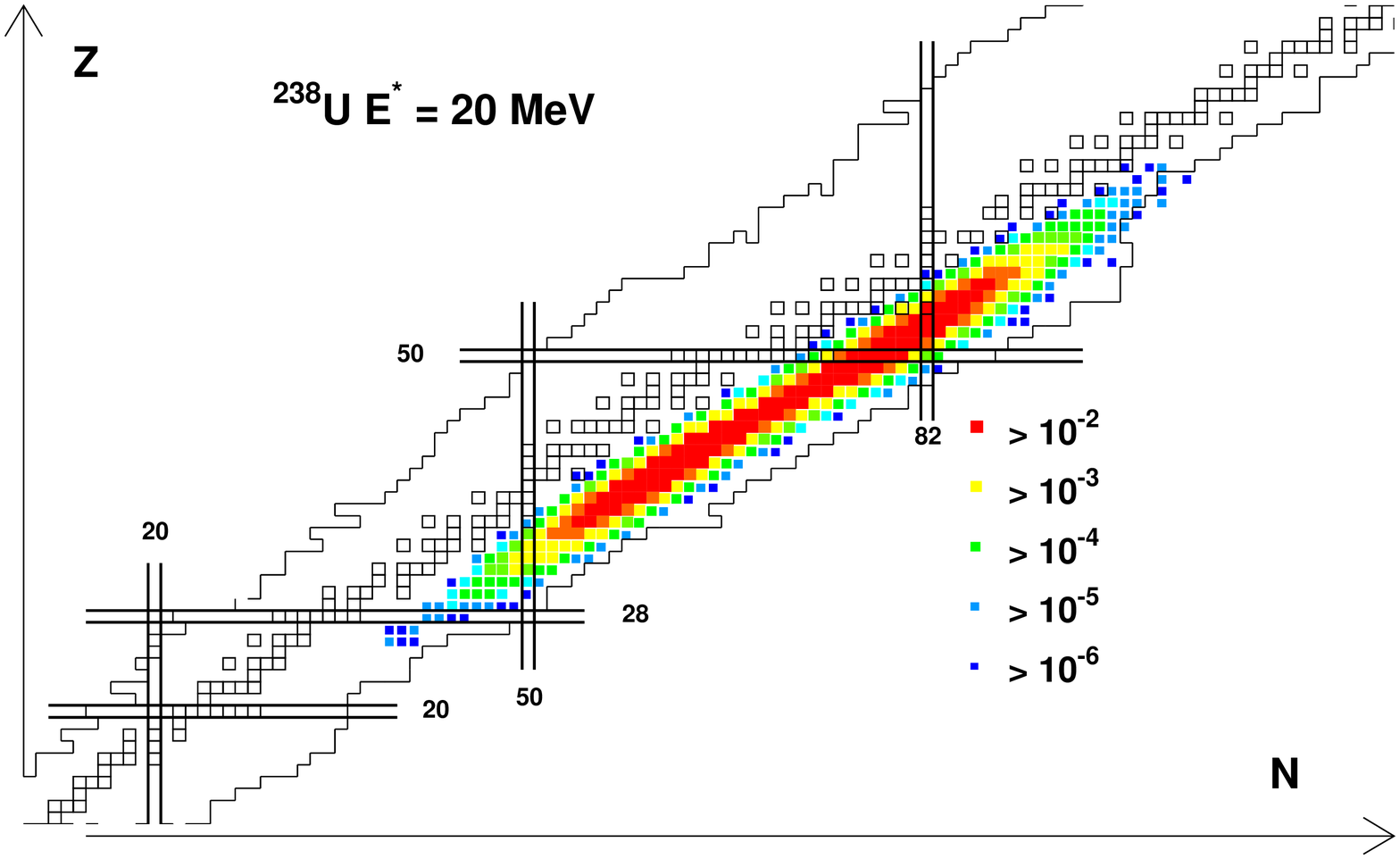,width=7.0cm}
    \epsfig{file=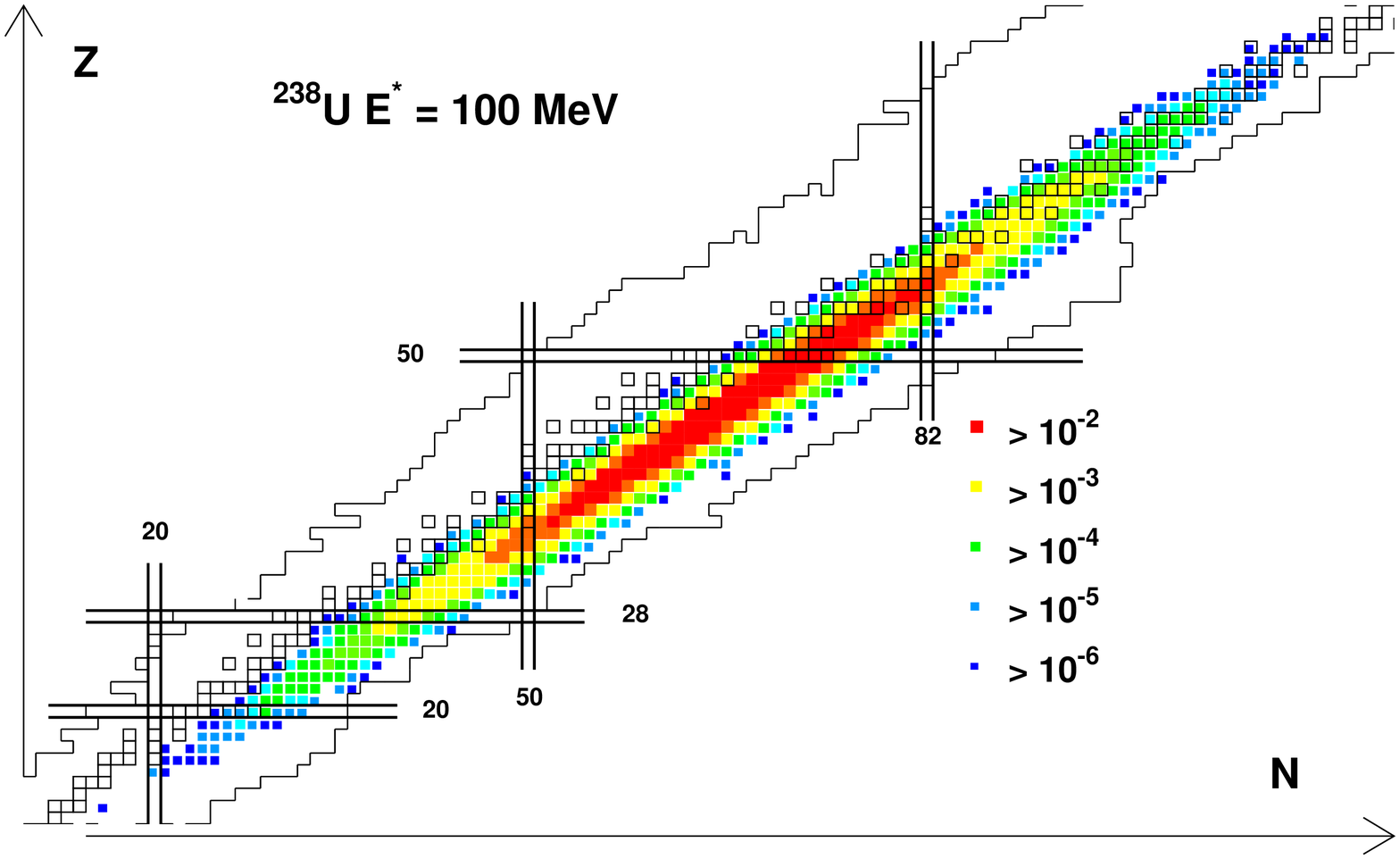,width=7.0cm}
\hspace*{-0.5cm}
\caption{\em Residual nuclei produced after the fission of $^{238}$U at different excitation energies.}
\label{fig_4}
\end{figure}

\section{Production of medium-mass neutron-rich nuclei}

\subsection{Fission}

Fission has largely been used to produced medium-mass neutron-rich nuclei up to the present limits of the chart of the nuclides \cite{ber97}. The isotopic distribution of residues produced in fission can be understood in terms of the potential governing this process. The Coulomb term of the nuclear potential is responsible for the neutron excess of the stable fissile nuclei leading to fission residues with an even larger neutron excess with respect to the valley of beta stability. However, the asymmetry term preserves the same N/Z of the fissioning nucleus in the fission residues. Shell effects and temperature induce a polarization effect which allow to produce even more neutron-rich residues.

In order to investigate the fluctuations in N/Z and mass asymmetry induced by the temperature of the fissioning system we have performed several simulations with the fission code of Ref.\cite{ben98}. In figure \ref{fig_4} we represent on top of a chart of the nuclides the distributions of residues after the fission of $^{238}$U at different excitation energies. As can be seen in the figure, when increasing the excitation energy, shell effects (double humped distribution) disappear, and the fluctuations in mass asymmetry and N/Z increase, populating a larger variety of neutron-rich residues. However, at high excitation energies neutron evaporation becomes more important, and the residue distribution moves to the neutron-deficient side. One can define an optimum excitation energy of the fissioning system around 50 MeV to produce the largest variety of neutron-rich nuclei.

\begin{figure}[ht]
\begin{center}
\epsfig{file=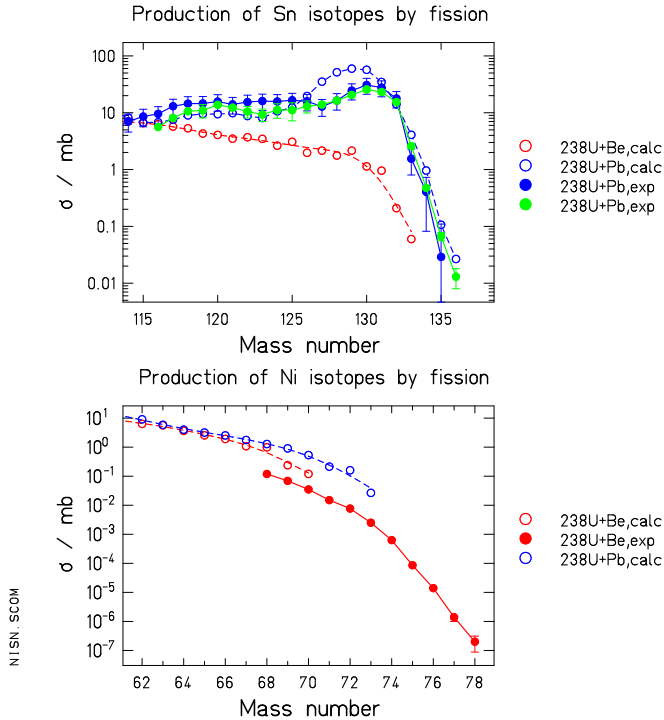,width=9.0cm} 
\end{center}  
\caption{Production cross sections of tin and nickel isotopes produced in the fission of $^{238}$U induced by different projectiles} 
\label{fig_5}
\end{figure}

Once we know the optimum conditions to produced neutron-rich nuclei in fission reactions, the final production rates will be defined by the evolution of the production cross sections. In Fig.\ref{fig_5}, we report the production cross sections of several neutron-rich tin and nickel isotopes produced in the fission of $^{238}$U projectiles on Be\cite{ber97}, Pd\cite{enq99} and H$_2$\cite{ber03}. These data show a dramatic decrease of the production cross section of around one order of magnitude per additional neutron. Consequently, one can not expect a spectacular expansion of the chart of the nuclides in the region of medium-mass neutron-rich isotopes by using fission reactions with the future rare-beam facilities.

\subsection{Two-step scenario: fission + cold fragmentation}
\begin{figure}[ht]
\begin{center}
\epsfig{file=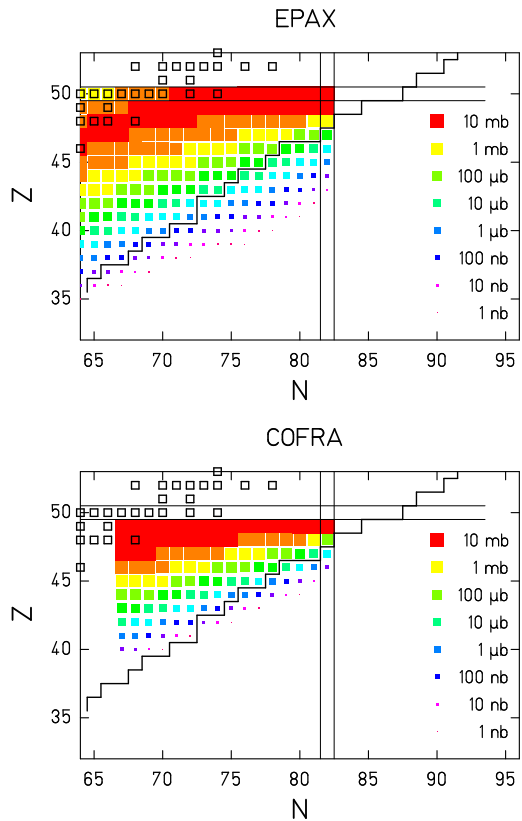,width=8.0cm}
\end{center} 
\caption{Predicted production cross sections of residues in the fragmentation of $^{132}$Sn at 1 A GeV on a beryllium target with the codes EPAX$^{10}$ (upper figure) and COFRA$^2$ (lower figure). The color scale represents the production cross section.} 
\label{fig_6}
\end{figure}

\begin{figure}[ht]
\centerline{\epsfig{file=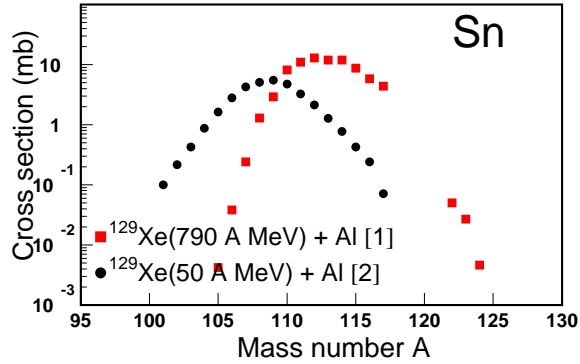,width=8cm}}   
\caption{Measured production cross sections of tin isotopes in the reaction $^{129}$Xe on aluminum at 790 A MeV (squares)$^{11}$ and 50 A MeV (dots)$^{12}$.} 
\label{fig_7}
\end{figure}

Recently it has been proposed to use both fission and cold-fragmentation reactions, in a two-step reaction scheme\cite{hel03}, in order to overcome the limitations of fission to produce extremely neutron-rich nuclei in the medium-mass region. However, nowadays it is difficult to make reliable predictions of the final production rates using this idea. As shown in Fig. \ref{fig_6}, present fragmentation codes show clear discrepancies in the predicted production rates when neutron-rich projectiles are used. In addition, the energy at which the fragmentation stage takes place plays a major role as shown in Fig. \ref{fig_7}. In this figure we report the measured production cross sections of different tin isotopes in the reaction $^{129}$Xe on aluminum at 790 A MeV \cite{rei98} and 50 A MeV \cite{han95}. According to these data the higher energies enhance the production of neutron-rich residues.

\begin{figure}[ht]
\epsfig{file=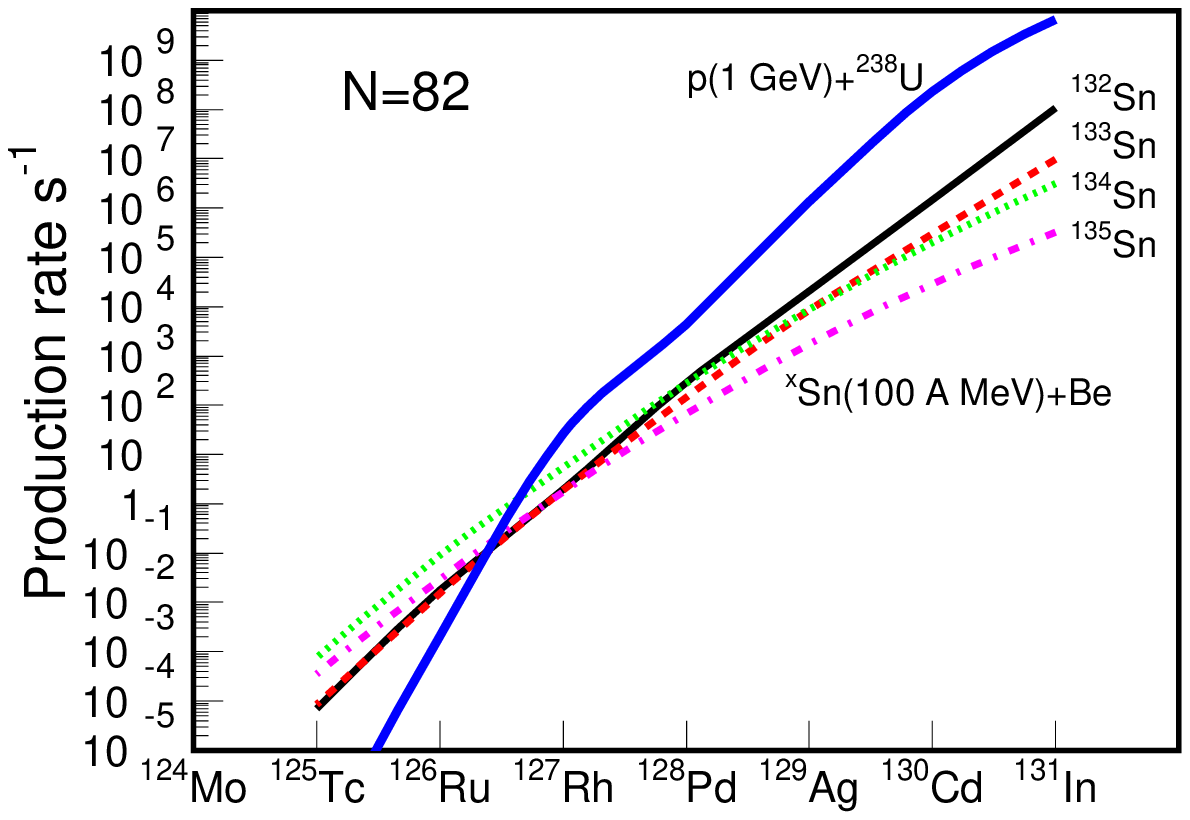,width=7.0cm}
\epsfig{file=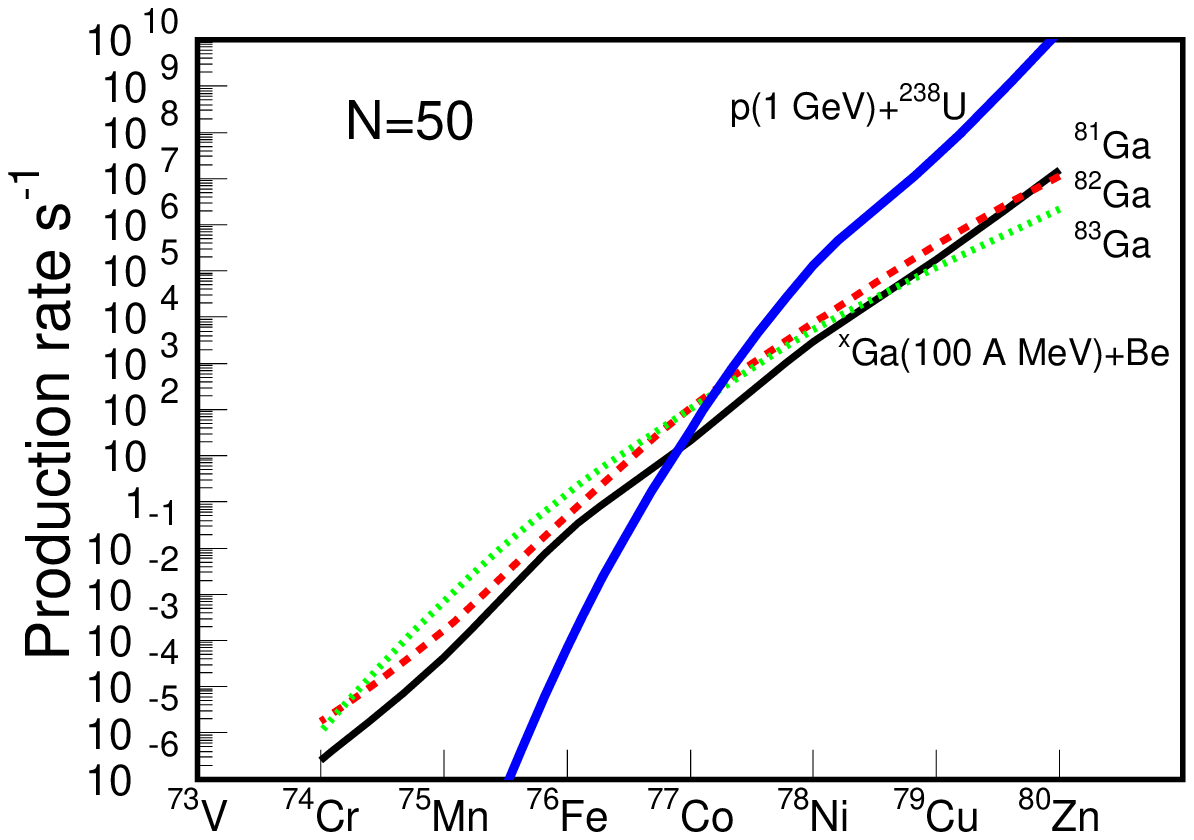,width=7.0cm}
\vspace*{-0.3cm}
\caption{Production of neutron-rich isotopes along the neutron shells N=50 and N=82 after the fission of $^{238}$U induced by 1 GeV protons (thick lines), and after the cold fragmentation in a beryllim target of different neutron-rich gallium and tin isotopes produced in by fission (thin lines).} 
\label{fig_8}
\end{figure}

Nevertheless, we used the cold-fragmentation code COFRA, which is one of the more reliable codes to predict the production of fragmentation residues from neutron-rich projectiles to estimate the production rates in a two-step reaction scheme at energies above 100 A MeV. In these calculations, the primary production cross sections in the fission step were taken from measured data in the reactions $^{238}$U(1 A GeV)+p \cite{ber03} and $^{238}$U(750 A MeV)+Be \cite{ber97}. The most representative results of these calculations are shown in figure \ref{fig_8}. In this figure we represent the production of different neutron-rich isotopes along the neutron shells N=50 and N=82. The thick line represents the direct production in the fission of $^{238}$U induced by 1 GeV protons, while the thin lines correspond to the two-step production after the cold fragmentation in a Be target of different Ga and Sn isotopes produced by fission. As can be observed in both pictures, the two-step scheme is competitive with the direct production only for the largest neutron excess. 

However, if one considers extraction, ionization and re-acceleration efficiencies, the situation could change. In this realistic scenario the two-step scheme could be used to produce by fission an abundant long-lived neutron-rich nuclei like $^{132}$Sn and fragmenting it to produce those neutron-rich isotopes that have low extraction efficiencies.

\section{Conclusion}
In this paper we have reviewed the two reaction mechanisms best suited to extend the present limits of the chart of the nuclides in the neutron-rich side. Recent experiments have shown that the cold-fragmentation process constitutes the appropriate tool to produce heavy neutron-rich residues. In the intermediate-mass region it is well established that fission allows to produce abundantly moderately neutron-rich isotopes. However, it will be difficult to enlarge the present limits of the chart of the nuclides using this reaction mechanism. An alternative would be to use a two-step scenario where the cold fragmentation of neutron-rich isotopes produced by fission is foreseen. Although this new scenario is only competitive with fission for isotopes with a large neutron excess, it would be an optimum solution to produce neutron-rich isotopes with low extraction efficiencies.


\begin{thebibliography}{0}

\bibitem{ben99} J. Benlliure et al., {\it Nucl. Phys.} {\bf A660}, 
87 (1999).

\bibitem{jon98} M. de Jong et al., {\it Nucl. Phys.} {\bf A628}, 479 (1998).

\bibitem{sch92} K.-H. Schmidt et al., {\it Nucl. Phys.} {\bf A542}, 699 (1992).

\bibitem{ber97} M. Bernas et al., {\it Phys. Lett.} {\bf B415}, 111 (1997).

\bibitem{ben98} J. Benlliure et al., {\it Nucl. Phys.} {\bf A628}, 458 (1998).

\bibitem{enq99} T. Enqvist et al., {\it Nucl. Phys.} {\bf A658}, 47 (1999).

\bibitem{ber03} M. Bernas et al., in preparation

\bibitem{hel03} K. Helariutta et al., submitted to {\it Eur. Phys. J}. {\bf A} 

\bibitem{sum00} K. S\"ummerer, B. Blank, {\it Phys. Rev.} {\bf C61}, 034607 (2000).

\bibitem{rei98} J. Reinhold et al., {\it Phys. Rev.} {\bf C58}, 247 (1998)

\bibitem{han95} K.A. Hanold et al., {\it Phys. Rev.} {\bf C52}, 52 (1995)

\end{thebibliography}
\end{document}